\documentstyle[preprint,pra,aps]{revtex}  

\newtheorem{prop}{Proposition}

\relax 
\begin{document}
\draft

\title{Formal context for cryptographic models}  

\author{John M. Myers}

\address{Gordon McKay Laboratory, Division of Engineering and Applied
Sciences,\\ Harvard University, Cambridge, MA 02138\\[-12pt]}

\author{and\\[-12pt]}

\author{F. Hadi Madjid}

\address{82 Powers Road, Concord, MA 01742} 

\maketitle 

\vspace*{-.5\baselineskip}
\begin{abstract}\indent To clarify what is involved in linking models to
instruments, we adapt quantum mechanics to define models that display
explicitly the points at which they can be linked to statistics of results
of the use of instruments.  Extending an earlier proof that linking models
to instruments takes guesswork, we show: Any model of cryptographic
instruments can be {\em enveloped}, nonuniquely, by another model that
expresses conditions of instruments that must be met if the first
model is to fit a set of measured outcomes.  As a result, model
$\alpha$ of key distribution can be enveloped in various ways to
reveal alternative models that Eve can try to implement, in conflict
with model $\alpha$ and its promise of security.  A different
enveloping model can help Alice and Bob by expressing necessities of
synchronization that they manipulate to improve their detection of
eavesdropping.  Finally we show that models based on pre-quantum
physics are also open to envelopment.
\end{abstract}

\vspace*{\baselineskip}

\narrowtext
\section{Introduction}

A designer Diana and the users Alice and Bob of cryptographic
transmitting and receiving instruments, as well as the eavesdropper
Eve, all employ various equations to model how results of the use of
these instruments depend on what the participants do.  Between a model
as a set of equations and instruments made of glass and silicon there
is a great divide.  In choosing a model to analyze instruments or to
be employed in a feedback loop where the model helps to operate
instruments, one makes a link across this divide.  While one can
interpret measured results as refuting some candidate models, we
recently proved that neither they nor logic can uniquely determine a
quantum model: linking a model to instruments requires something
beyond logic and measured data, something well named by the word {\em
guess} \cite{ams}.

Proofs of the security of quantum key distribution invoke inner
products of quantum state vectors, and these depend on the model chosen.
Here we prove that any given set of outcomes from a transmitter and
receiver used to distribute a key can be fitted by many
quantum-mechanical models which differ greatly among themselves in
their inner products and hence in their implications for the security
of a key.  On one hand, this encourages Eve to invent snooping
instruments even though she knows Alice and Bob have a proof of
security, and, on the other hand, our findings encourage discovery and
repair of ``hidden security loopholes'' \cite{lo}
arising because their transmitting and receiving instruments ``violate
... assumptions [that underlie their model] in ways not immediately
apparent to Alice and Bob'' \cite{slutskyPRA}.

To underpin an examination of the linking of models to instruments, in
Section II we adapt quantum mechanics to define models that display
explicitly the points at which they can be linked to statistics of results
of the use of instruments.  The models to be introduced express ``what
the participants do'' in terms of {\em commands} sent to the
instruments {\em via} Classical, digital Process-control Computers
(CPC's) that control them and that also record results from them; we
call these CPC-oriented models. Section III extends an earlier proof
that linking CPC-oriented models to instruments takes guesswork: For
any quantum-mechanical model of transmitting and receiving instruments
there is another model (not unique) that expresses constraints in
using the instruments that must be met if the first model is to fit a
set of measured outcomes.  We say the first model is {\em enveloped}
by the second. 

In Section IV we prove that for any quantum
model $\alpha$ of key distribution, there exists an enveloping model
$\beta$ that matches $\alpha$ with respect to measurements
contemplated in $\alpha$ but that has smaller inner products and
allows for other measurements, which, if Eve can implement them, allow
undetected eavesdropping, in conflict with model $\alpha$ and its
promise of security.  For this reason, no proof can relieve Alice and
Bob of the burden of making judgments about what models to link to
their instruments, something implicit in \cite{lo,slutskyPRA}, but
here made vivid.

They can, however, put the same burden of judgment on Eve, for
she too must use models.  In Section V model $\alpha$ is enveloped by
another model that expresses necessities of synchronization that Alice
and Bob can manipulate to improve their detection of eavesdropping.
In Section VI we indicate how models based on pre-quantum physics are
also open to envelopment.

In summary, we find that instruments modeled are used in a context of
circumstances and intentions which no model can fully describe.  In
creating an enveloping model, one formally expresses (rightly or
wrongly) some hitherto unexpressed feature of this context.  As will
be proved, there is no end of opportunities to assert features of
context, because any enveloping model can in turn be enveloped.

\section{Linking instruments to models}

The central issue is the linking of uses of instruments to models.  By
{\em model} we mean a set of equations written in mathematical
language, primarily quantum mechanics, with the intent of predicting
statistics of the results of using instruments (such as transmitters
and receivers made of silicon and glass fibers {\em etc.}).  Some of
the equations of the set act as a set of assumptions from which the
rest of the equations can be derived.  Quantum mechanics provides a
mathematical language in which to write down a wide variety of models,
constrained by a grammar of logical constraints, so within a model
conclusions can be proved to follow from assumptions.  Because
different sets of assumptions generate different quantum-mechanical
models, quantum mechanics is a language, as distinct from a particular
model written in that language; it has more room in it for diverse
models that accord with any given set of experimental results than has
been appreciated.

Although, as we shall see, instruments cannot be discussed
independently of models, we separate them as best we can by the trick
supposing the instruments are operated via digital computers.  This
will allow us to express ``how the instruments are used'' in terms of
{\em commands} sent to the instruments by a Classical, digital
Process-control Computer (CPC) that controls them and that also
records results from them \cite{ams}.  Instruments swallow commands
and give back recordable results.

As discussed in \cite{ams}, parsing a stream of data from instruments
into a sequence of measurement occurrences, each with a
quantum-mechanical {\em outcome}, cannot be determined from the data
alone, but takes extra hypotheses, indeed a kind of stripped-down
model, determined in part by guesswork, which we call a {\em parsing
rule}.  Parsing requires guesswork, both to assert the statistical
independence of one segment of data from another, and to select
criteria by which to weed out artifacts in the data attributed to
instrumental imperfections, such as false and missed detections.
Using a parsing rule, one parses a stream of data from the
instruments into a sequence of measurement occurrences, assumed
statistically independent, and one formats the data segment for each
occurrence into (1) how the instruments were configured and (2) an
outcome from the instruments.  The parsing rule makes no statement
about values of probabilities of outcomes, but does assert that such
values exist; it provides a range of outcomes that are possible to
record as well as set of possible command sequences.  In this way it
limits the models that can be tested by measured data that it parses.

To view the linking of instruments to models we postulate an analytic
frame in which (a) instruments write via some parsing rule what we and
other scientists interpret as numerical outcomes, (b) they write these
numbers in memories of CPC's, and (c) CPC's send commands to some or
all of the instruments.  We view each set-up of instruments in terms
of records in CPC memories of commands sent to the instruments by
CPC's and of outcomes from the instruments.  Notice that we make no
assumption that an instrument works as the manufacturer says it does,
nor that it works the way any model compatible with the parsing rule
says it does, nor that it functions statistically the same on
Tuesday as it does on Monday.  While such assumptions are made in
the models to be discussed, the analytic frame provides room to
consider cases in which the instruments write numbers that conflict
with any or all models.

We define a CPC-oriented model of a set-up of instruments to be a
model that expresses conditional probabilities of outcomes given
commands to the instruments.  For instance, a model $\alpha$ to be
introduced in Section IV will express a conditional probability of
outcome $j$ given a command $b_A$ from Alice's CPC to her transmitter
and a command $b_E$ from Eve's CPC to her eavesdropping receiver,
written $\Pr_\alpha(j|b_A,b_E)$. The subscript marks it as an
assertion within model $\alpha$, leaving room to consider a different
model $\beta$ that asserts a different numerical value
$\Pr_\beta(j|b_A,b_E) \neq \Pr_\alpha(j|b_A,b_E)$.\footnote{This
extends our earlier analysis of instruments controlled by a single CPC
programmed according to some model \cite{ams,spie} to deal with a setup
of instruments controlled by several CPC's.} 

The same CPC's that control the instruments house in their memories
CPC-oriented models and programs designed using them.  These models
and model-derived programs are used off-line to simulate the
instruments; they are used on-line, not to simulate the instruments,
but to help operate them, for example in a feedback loop of Bob's
receiver, as discussed in Section~V. By considering both the
instruments and the models as they are reflected in files of a CPC,
we conceptually separate (as well as possible) these CPC-oriented
models from the instruments modeled while allowing for interaction 
between models and instruments.

Like any set of equations, a CPC-oriented model can be copied, so
copies of the same model can be used concurrently in different places
for the same or different purposes.  What can be done with a model or
a program depends on where it is, for example on whether it is written
in Alice's CPC or in Eve's. Because the models used in programming one
CPC need not be the same as those used in programming another, several
CPC's controlling interacting instruments can work from different
models concurrently.  Where and when and how a CPC-oriented model is
used is traceable in the execution sequences of the CPC's in which
copies of the model are housed, so the CPC frame allows analysis of
various of CPC-oriented models and model-derived programs used to
operate instruments that interact. Some or all of the instruments can be
modeled by more than one model, and one model can conflict with
another.  Some models model other models: a component of Eve's model
can be her model $\alpha'$ of Alice and Bob's model $\alpha$; this
tells her (rightly or wrongly) how Alice and Bob, using their model
$\alpha$, will decide on their security, distinct from how Eve decides
using her model $\beta$.  Conversely, Alice and Bob's model $\alpha$
contains as a component a model $\beta'$, their model of Eve's model
$\beta$.  There is no necessary stopping place in modeling models.

If a model $\alpha$ invokes all the assumptions of a model $\beta$ and
possibly more, we say model $\alpha$ {\em specializes} model $\beta$,
or that model $\beta$ {\em generalizes} model $\alpha$ (meaning it has
fewer assumptions).  This is the first of several types of relations
among models that will be used to express interactions between the
invention and the modeling of transmitting and receiving instruments
used in cryptographic key distribution.

\section{Models of communication}

Ignoring eavesdropping for the moment, we focus on Alice communicating
to Bob, as described quantum mechanically.  Consider Alice
transmitting $m$ quantum bits of raw data to Bob, with Alice using one
CPC to control her transmitter and Bob using another to control his
receiver.  They want to jointly implement a CPC-oriented model
$\gamma$ of quantum communication, which says at each of a sequence of
$m$ occurrences, Alice causes her CPC to command the preparation of a
state vector, choosing ${\bf u}$ for 0 or ${\bf v}$ for 1 at random.
Her CPC records her choices of 0 or 1.  Bob's receiver has, say, light
detectors, one interpreted in model $\gamma$ as detecting ${\bf u}$ to
indicate Alice's choice of 0 and another detecting ${\bf v}$ to
indicate Alice's transmission of 1.  Bob's CPC records the decisions
of his receiver, described as deciding on 0 or 1 or, if neither
detector fires, `inconclusive' \cite{B92,ekert94}.

Model $\gamma$ is composed of: 
\begin{enumerate} 
\item a set $A_\gamma$ of command strings that Alice's CPC can send to
generate states, here just the set $\{0,1\}$; a set $B_\gamma$ of
commands that Bob's CPC can send to his receiver, which in this case
is empty; 
\item a Hilbert space ${\cal H}_\gamma = {\bf C}^2$ ({\em i.e.},
the vector space of complex dimension 2);
\item a function for states as functions of commands (here only
Alice's commands), $|v_\gamma \rangle : A_\gamma \rightarrow
{\cal H}_\gamma$ such that $|v_\gamma(0) \rangle = {\bf u}$ and
$|v_\gamma(1) \rangle = {\bf v}$;
\item a set of possible outcomes of Bob's measurement, indexed by $j$
ranging over natural numbers or some subset of natural numbers, here 0
for $|v_\gamma(0) \rangle$, 1 for $|v_\gamma(1) \rangle $, and 2 for
`inconclusive';
\item a function from Bob's commands to positive operator valued
measures (POVM's) on ${\cal H}_\gamma$, here simplified to the
single POVM $M_\gamma$ consisting of a set of detection operators
$M_\gamma(j)$ with
\begin{enumerate} 
\item[] $\sum_j M_\gamma(j) = {\bf 1}$, 
\item[] $(\forall j) M_\gamma(j) \geq 0$ and $M_\gamma(j) =
M_\gamma(j)^{\dag}$.
\end{enumerate}
\end{enumerate} Model $\gamma$ asserts the probability of outcome $j$
given a command $b_A \in A_\gamma$ for state preparation to be
\begin{equation} \mbox{$\Pr_\alpha(j|b_A)$} = \langle v_\alpha(b_A) |
M_\gamma(j)| v(b_A) \rangle. \end{equation}

In relating model $\gamma$ to results in his CPC, Bob thinks
of his CPC as recording detection results of Alice's $m$-bit
transmission in a sequence of $m$ memory segments, each of which can
hold two bits, coded 00 for Alice's `0', 01 for Alice's `1', and 10
for `inconclusive'.  We will refer to these two-bit memory segments
in connection with timing, to which we now turn.

\subsection{Need for synchronization}

Model $\gamma$ is an armchair view of Bob's receiver that lacks the
detail necessary to design it.  To design a receiver that works
according to model $\gamma$, Diana must provide for synchronizing it
to Alice's transmitter within some allowed leeway.  For this Diana
envelops model $\gamma$ with a more detailed model $\delta$ that
expresses the conditions of synchronization that must be maintained
between Alice's transmitter and Bob's receiver if model $\gamma$ is to
accord the records of Alice's commands and Bob's detections.  Diana
provides for Bob's receiver to meet these conditions by adjusting the
rate of Bob's clock in response to measured results interpreted in
model $\delta$.  She designs this feedback loop by choosing a
classical-control model $\epsilon$, to be discussed shortly.  Without
the gaps in synchronization defined by model $\delta$ and their
containment within the allowed leeway in accord with model $\epsilon$,
Bob's CPC, driven by its clock, would mistime its routing of a
detector signal to the $k$-th memory segment, resulting in an
erroneous record.\footnote{Although for a short transmission line of a
fixed delay, Bob and Alice can use the same clock to drive their CPC's
synchronously, but for variable delay, {\em e.g.} if Bob is in motion,
he needs his own clock, independently adjustable \cite{hj84}. And even
where the single-clock design works, Bob's receiver must adjust its
phase.}

To express the effect on reception of the drift of the clock of Bob's
CPC relative to Alice's clock, Diana (having learned from Einstein)
defines synchrony in terms of measurements made of Alice's signal
arriving at Bob's clock; however, in using a quantum model of the
measurement of that signal, she must cope with quantum indeterminacy,
which limits Bob's receiver's knowledge of arrival times to what it
can deduce via Bayes rule from outcomes.  To produce suitable
outcomes, Diana invents a model $\delta$, which has a Hilbert space
${\cal H}_\delta$, of dimension higher than that of
${\cal H}_\gamma$, along with states $|v_\delta(b_A,s) \rangle$
that are functions not only of Alice's commands in $A_\delta =
A_\gamma$, but also of a {\em skew} $s$ of Bob's clock relative to an
imagined ideally synchronized clock.  Diana designs Bob's receiver to
measure the $k$-th signal from Alice when the clock of Bob's CPC reads
$t_k$.  When we imagine Bob's clock reads $t_k$ as the ideal clock
reads $t_k - s_k$, we say Bob's clock is fast by a skew $s_k$.  The
state measured by Bob's receiver when his clock reads $t_k$ is then
$|v_\delta(b_A,s_k)\rangle = U_\delta (-s_k)|v_\delta(b_A,0) \rangle$,
where $U_\delta$ is a unitary-operator-valued function of $s_k$ by
which Diana expresses skew.  

In order to allow different possible outcomes for different values of
the skew $s_k$ at the reception of the $k$-th of the $m$ signals from
Alice, model $\delta$ must assume more possible outcomes than the 0,
1, and `inconclusive' of model $\gamma$, so the POVM $M_\delta$ has
more than the three detection operators of $M_\gamma$.  When
restricted to skews $s_k$ of magnitude smaller than some allowed bound
$s_0$, model $\delta$ projects onto model $\gamma$ as follows:
\begin{enumerate} \item $|v_\delta(b,s_k)\rangle \mapsto |v_\gamma(b)
\rangle$, and \item the detection operators of $M_\delta$ partition
into three sets, such that the sum of operators for each set maps to a
single detection operator $M_\gamma(j)$ with respect to probabilities
of detection of $j$. \end{enumerate} But outcomes in model $\delta$
tell more than these projections.  At each signal reception, Bob's
receiver records in his CPC not only a decision among 0, 1, and
`inconclusive' but also finer distinction from which his CPC estimates
its clock skew (via Bayes rule and a prior probability distribution
that Diana assumes for skew).  In order to record the outcomes that
help estimate skew and guide clock-rate adjustment, Bob's receiver,
designed using model $\delta$, needs a memory segment for the $k$-th
reception of more than two bits.  Hence, the record previously discussed in
connection with model $\gamma$ is extracted from a larger record
required by model $\delta$.\footnote{Must there exist a
quantum mechanical model that accords with experimental results of
measurements of a skew-dependent state?  Yes, because, any digital
record can be interpreted (nonuniquely) as a record of quantum
outcomes, and for any set of outcomes with their relative frequencies
as functions of commands, many quantum mechanical models have
probabilities that exactly fit \cite{ams}.}

To contain skews within the tolerable bound $s_0$, Diana chooses a
classical model $\epsilon$ by which to design a program that, when
executed by Bob's CPC, responds to estimated skews by sending a
command from the CPC to set a `faster-slower' lever on the clock that
drives that CPC; the command is a value of a {\em control function}
$F_\epsilon$ that takes as its argument a computer file consisting of
skews calculated from recently recorded detection results and recently
issued commands to the clock itself.  Although the quantum state to be
controlled has a history that Bob's CPC can only estimate via Bayes
rule from outcomes, the design of a control function $F_\epsilon$ is
within the discipline of classical feedback design.\footnote{For
discussion of Bayes rule in a non-quantum context of control, see
\cite{jazwinski}.}  If model $\delta$ is implemented and if model
$\epsilon$ succeeds in generating steering commands that are adequate,
the skews are held within the bound $\pm s_0$ so that Bob's detection
results fulfill the intention of model $\gamma$ and, additionally,
allow his CPC to make skew estimates necessary to guide clock
adjustment.  

\vskip\abovedisplayskip
\noindent{\bf Remark 1}: Models, such as
$\gamma$ and $\delta$, express desires and obstacles more flexibly
than do {\em inputs} used for this purpose in control theory
\cite{jazwinski,stefani}.  Alice expresses what she wants by choosing
model $\gamma$ altogether, not just by an `input' of 0 or 1 to her
transmitter.  Because Diana wants Alice and
Bob's instruments to work in accord with Alice's model $\gamma$, in
spite of the obstacle of clock drift, she chooses models $\delta$ and
its classical companion, model $\epsilon$. 

\vskip\abovedisplayskip
\noindent{\bf Remark 2}: The number of bits that
arrive at Bob's receiver is model-dependent: whether a detection result
for a signal is seen as two bits (ignoring skew) or as more bits (allowing
for skew) depends on whether the record of the signal detection is
interpreted using model $\gamma$ or model $\delta$.

\vskip\abovedisplayskip
Recall the freedom always present in quantum
mechanical modeling to shift the boundary between the `system' modeled and
the measuring instrument, for instance by counting more of the measuring
instrument as part of the system \cite{peres}.  In view of this freedom, we
conclude: 

\vskip\abovedisplayskip
\noindent{\bf Remark 3}: Every quantum mechanical
model is contingent in the sense that it is projected onto by a
restriction of an enveloping model that shows other possibilities.

\section{Models of vulnerability to eavesdropping}

Widely discussed quantum-mechanical models of key distribution assert
a nonzero inner product between quantum state vectors that Alice
communicates to Bob, with the consequence that eavesdropping almost
always leaves tracks in the form of errors that Bob and Alice can
detect.  If Alice's transmitter, Bob's receiver, and Eve's snooping
instruments can be counted on to work in accord with any of these
models, then Alice can send Bob a key secure against undetected
eavesdropping.  The models can all be translated into CPC-oriented
models to make visible the points at which they can be linked to
results of the use of instruments, and it is to the credit of some of
these models that relative frequencies of experimental results accord
reasonably well with conditional probabilities of outcomes derived
from the states and operators posited by the models.  But we are
sloppy if we forget that quantum states are terms in models, rather
than model-independent features of instruments.

In linking a CPC-oriented model $\alpha$ to instruments, one
identifies commands in model $\alpha$ with commands sent from CPC's to the
instruments, for example commands $b_A$, $b_B$, and $b_E$ from CPC's
controlled by Alice, Bob, and Eve, respectively; one also parses
results of the use of instruments in response to commands as
quantum-mechanical outcomes, so that one can compare relative
frequencies of these results to the conditional probabilities asserted
by model $\alpha$, e.g. $\Pr_\alpha(j|b_A,b_B,b_E)$ as the conditional
probability of a quantum outcome $j$ given commands $b_A$, $b_B$, and
$b_E$.  (The outcome $j$ can be seen as several fragments, for example
one for Bob and one for Eve, allowing for analysis of mutual
information between Eve and Bob, etc.) It is to be noticed that this
procedure sets up a divide that runs through the CPC between state
vectors as terms in models, on one side, and on the other side the
commands to and results from instruments.  A large part of the story
told here amounts to noticing this divide.

Given a CPC-oriented model $\alpha$ of quantum key distribution that
shows Alice and Bob to be secure against eavesdropping, one can
envelop model $\alpha$ in a model $\beta$ that introduces a range of
conditions; under some conditions model $\beta$ projects to model
$\alpha$, agreeing with it, while under other conditions model $\beta$
leads to drastically different conclusions in conflict with those of
model $\alpha$.  Among these are conditions under which Eve can learn
the key without leaving tracks that Alice and Bob can detect.  This
envelopment is possible because model $\beta$ can invoke states and
their inner products that differ from those of model $\alpha$ while
still agreeing with model $\alpha$ with respect to probabilities
of outcomes for commands considered
in model $\alpha$.

For example, we envelop model $\alpha$ with a model $\beta$ expressing
conditions in which Alice's transmitter leaks light into a channel
accessible to Eve, but that is unknown to Alice and Bob (and is not
expressed in model $\alpha$ \cite{slutskyPRA}).  There are two cases
to consider, corresponding to two types of models.  Deferring models
of Eve's use of a probe, we start with the simpler case of a model
that segments the transmission of signals from Alice to Bob into (1)
Alice's transmission to Eve, followed by (2) Eve's transmission to
Bob.  For such segmented transmission, suppose model $\alpha$ assumes
that (1) Alice chooses commands from a set $A_\alpha = \{0,1\}$, with
command $b_A$ generating a state vector $|v_\alpha(b_A) \rangle \in
{\cal H}_\alpha$, and (2) Eve commands her listening instruments
with a command $b_E \in E_\alpha$ to make a measurement expressed by a
POVM $M_\alpha(b_E)$ which has a detection operator
$M_\alpha(b_E;j_E)$ acting on ${\cal H}_\alpha$, associated with
outcome $j_E$.  Model $\alpha$ implies that the conditional
probability of Eve obtaining the outcome $j_E$ given her command $b_E$
and Alice's command $b_A$ is
\begin{equation} \mbox{$\Pr_\alpha(j_E|b_A,b_E)$} =
\langle v_\alpha(b_A) | M_\alpha(b_E;j_E)| v_\alpha(b_A) \rangle. 
\label{eq:p_alpha}\end{equation}

\begin{prop} Given any such (segmented) model $\alpha$ with
inner pro\-duct $\langle v_\alpha(0)|v_\alpha(1) \rangle$ and given
any $0 \leq r < 1$, there is a model $\beta$ that gives the same
conditional probabilities of Eve's outcomes for all her commands
belonging to $E_\alpha$, so
\begin{equation} (\forall
b_A \in A_\alpha, b_E \in E_\alpha) \mbox{$\Pr_\beta(j_E|b_A,b_E)$} =
\mbox{$\Pr_\alpha(j_E|b_A,b_E)$} \label{eq:same} \end{equation} while
\begin{equation} |\langle v_\beta(0)|v_\beta(1)\rangle | = r
|\langle v_\alpha(0)|v_\alpha(1)\rangle | . \label{eq:inner}
\end{equation} \end{prop}

\noindent {\em Proof}: Motivated by the idea that, unknown to Alice,
her transmitter signal might generate an additional ``leakage'' into
an unintended spurious channel that Eve reads, we construct the following
enveloping model $\beta$ which assumes:
\begin{enumerate} \item the same set of
commands for Alice, so $A_\beta = A_\alpha$,
\item a larger Hilbert space ${\cal H}_\beta =
{\cal H}_{\rm leak} \otimes {\cal H}_\alpha$ in which Alice
produces vectors $|v_\beta(b_A) \rangle = |w_\beta(b_A) \rangle \otimes 
|v_\alpha(b_A) \rangle$, with $|w_\beta(b_A) \rangle \in
{\cal H}_{\rm leak}$; 
\item a larger set of commands for Eve,  $E_\beta = E_\alpha
\sqcup E_{\rm extra}$ (disjoint union);
\item a POVM-valued function of Eve's commands to her measuring
instruments, with detection operators
\begin{equation} M_\beta(b_E;j_E) = \left\{ \begin{array}{l}
{\bf 1}_{\rm leak} \otimes M_\alpha(b_E;j_E)
\mbox{ for all } b_E \in E_\alpha,\\
\mbox{Eve's choice of POVM to distinguish } |v_\beta(0) \rangle\\ 
\mbox{from } |v_\beta(1) \rangle \mbox{ if } b_E  \in E_{\rm
extra}. \end{array} \right. \end{equation} \end{enumerate} 
According to model $\beta$, if Eve chooses any measurement command of
$E_\alpha$, Eq.\ (\ref{eq:p_alpha}) holds.  But model $\beta$ speaks
not of the vectors $|v_\alpha(b_A) \rangle$ but of other vectors
having an inner product of magnitude
\begin{equation} |\langle v_\beta(0)|v_\beta(1)
\rangle | = |\langle w(0)|w(1) \rangle | 
|\langle v_\alpha(0)|v_\alpha(1) \rangle|. \label{eq:winner} 
\end{equation} 
The unit vectors $|w(0) \rangle$ and $|w(1) \rangle$ can be
specified at will, so that the factor $r
\stackrel{\rm def}{=} |\langle w(0) | w(1) \rangle |$ can be chosen
to be as small as one pleases.  $\Box$

If she can find and gain access to a channel carrying leakage states,
Eve implements a model $\beta$ with a value of $r < 1$, in which case
she uses an optimal POVM to distinguish Alice's 1's and 0's, with
fewer `inconclusives' than Alice and Bob think possible, and hence
with less impact on Bob's error rate.  If Eve can do this, she has more
information about the key for a given rate of Bob's errors than Alice
and Bob found possible when they bet on model $\alpha$, thus vitiating
Alice and Bob's attempt to distribute a key secure against undetected
eavesdropping.

Whether Eve can implement a measurement of leakage as called for in
model $\beta$ with $r < 1$ is unanswerable by modeling; it is a
question that requires work on ``the other side of the divide.''  The
point to be stressed is that the agreement between model $\alpha$ and
a set of measured results, no matter what results, is no logical
guarantee against Eve implementing model $\beta$ with a value of $r$
less than 1, or even a value of 0 which would give her the whole
key while causing no errors for Alice and Bob to detect.

\subsection{Models involving a defense function}

When noise in communications channels is recognized, privacy
amplification is necessary to distill a secure key \cite{bennett95}.
Arguments for the security of quantum key distribution with noisy
channels, summarized and refined in Refs.\
\cite{slutskyPRA,slutskyAO,brandt,brandt2}, center on a {\em defense
function}.  The existence of a defense function depends on a proof
(within some model) of a relation between Eve's maximum Renyi
information on whatever bits she directly or indirectly interrogates
and a positive contribution to Bob's error rate in receiving bits.

Defense functions have been analyzed for models of Eve's use of a
probe \cite{fuchs} and without restricting Alice's transmission to a
choice of only two state vectors.  In such a model $\alpha$, Alice
chooses one of several state vectors in one Hilbert space
${\cal H}_{{\rm sig},\alpha}$ while Eve generates a fixed vector in
a different Hilbert space ${\cal H}_{{\rm probe},\alpha}$, and the
tensor product of Alice's choice of state vector and Eve's fixed probe
vector evolves unitarily in an interaction, after which Eve and Bob
make measurements, Eve confined to the probe sector and Bob to the
signal sector.  Like segmented models, probe models relate Eve's
information to Bob's error rate in such a way that Bob's error rate
depends on inner products ascribed to the state vectors among which
Alice chooses; in particular if the inner products for distinct signal
vectors are all zero, Eve can learn everything without causing any
effect that Alice and Bob can detect.

The Appendix displays consequences of leakage of Alice's transmission
for models involving Eve's use of a probe: just as for models that
segment the transmission, the state vectors used to model Alice's
transmission are model-dependent, and so are their inner products.  To
see the consequence for defense functions, suppose that Alice and Bob
use model $\alpha$ which assumes that Alice chooses between state
vectors $|v_\alpha(0) \rangle$ and $|v_\alpha(1) \rangle$ with inner
product having a magnitude $S_\alpha = |\langle
v_\alpha(1)|v_\alpha(0) \rangle |$.  Assuming model $\alpha$, Alice
and Bob determine a defense function $t(n,e_T)$, as discussed in
\cite{slutskyAO}; in order to mark its dependence on model $\alpha$
and especially its dependence on the inner product of $S_\alpha$, we
write this as $t_\alpha(n,e_T,S_\alpha)$.  For any such model $\alpha$
and whatever the measured results with which it accords, we can show a
model $\beta$ that agrees with model $\alpha$ insofar as these results
are concerned, but disagrees with it about predictions of the
detectability of Eve's eavesdropping, because in place of the inner
product(s) of model $\alpha$, model $\beta$ has inner product(s)
smaller by our choice of $r$, for any $0 \leq r < 1$.  
\begin{prop} If
a model $\alpha$ asserts that Alice and Bob can distill a key that
is secure against measurements commanded by Eve from a set of commands
$E_\alpha$, then there exists another model $\beta$ that matches the
predictions of model $\alpha$ for the commands in $E_\alpha$ but that
makes additional commands available to Eve that make the key insecure.
\end{prop}

To prove this, one uses Proposition 3 of the Appendix that envelops
any model $\alpha$ with a model $\beta$ in which $S_\beta = r
S_\alpha$ with $r$ as small as one pleases.  The effect of making 
$S_\beta$ smaller than $S_\alpha$ is visible for the case of B92
models \cite{B92} in Figure 4 of \cite{slutskyAO}, where $S_\beta$
is denoted (in notation with which our notation regrettably clashes)
by $\sin 2\alpha$.  One sees that as $S_\beta$ gets smaller, $t_\beta$
gets bigger, so that at any fixed error rate, one can determine an $r$
for which model $\beta$ allows no distilled secure key.  For the BB84
model as discussed in \cite{slutskyAO}, the effect of $r \ll 1$ in an
enveloping model $\beta$ is to conflict with the BB84 model in such a
way as to increase $t$ and allow undetected eavesdropping.

Thus, just as for segmented eavesdropping discussed above, Eve can
try to implement a model $\beta$ which drastically increases what she
can learn for a given error rate.  Again, whether she can succeed in
implementing such a model is another question, on the other side of
the divide that runs through the CPC's between models and instruments.
No matter what measured results they stand on, Alice and Bob always
face a choice between a model $\alpha$ and an enveloping model $\beta$
that challenges the security asserted by model $\alpha$.  Because both
models make identical predictions about probabilities that connect with
the measured data, Alice and Bob face a choice that no combination of
logic and their fixed set of measured results can decide.  They must
make a judgment, or, to put it baldly, they must make a {\em guess}
and act on it \cite{ams}.

\section{Modulation of clock rate to improve security}

While Alice and Bob may view their need for guesswork and judgment as
bad news, they can put this need to good use if their system designer
Diana recognizes that Eve is in the same boat: she too must act on
guesswork.  Recognizing this, Diana can design a key distribution
system with features that make it harder for Eve to snoop. 

As discussed in Section III, to accord with model $\alpha$, any
receiver whether Bob's or Eve's, must maintain close synchrony with
Alice's transmission in order to function.  In both the segmented and
the probe cases discussed above, the models $\alpha$ and $\beta$ can
accord with measured results only if Bob's and Eve's receivers  work
in accord with enveloping models similar to model $\delta$ that
expresses clock skew contained within an allowed leeway.  Recall that
model $\delta$ describes a receiver as parsing its results for each of
Alice's bits into two parts, one indicating `0', `1', or
`inconclusive', the other indicating skew to be contained by adjusting
the faster-slower lever of Bob's clock, and, like Bob's receiver,
Eve's must do this to make eavesdropping measurements at times that
work.\footnote{In the segmented case, Bob, unaware, synchronizes his
receiver to Eve's re-transmission, even though he supposes he is
synchronizing with Alice's transmission.}

We suggest that Diana try to design Alice's transmitter and Bob's
receiver to make the parsing by Eve's receiver impossible without use
of prior information that Alice has also encoded, and that Bob has
better access to than does Eve.  The idea is for Alice's transmitter
to be timed by a clock whose rate is intentionally randomly varied
rapidly and over a wide range, and for Alice to encrypt indications of
coming rate variations in her transmission to Bob.  The eavesdropping
problem is different (and harder) for these rate variations than for
the key because they are more perishable.  Quantum-mechanical models
assert that the operation of the faster-slower lever on Eve's receiver
cannot be corrected {\em ex post}; that is, if she intercepts Alice's
signal and records it using a receiver clock unsynchronized to
Alice's transmission, there is no way to reconstruct from her record
what she would have received with a synchronized clock.

\section{Generalization}

Extending the proof in \cite{ams} that guesswork is necessary to the
linking of quantum models to results of instruments, we have introduced
the concept of enveloping models to prove that for any quantum-mechanical
model $\alpha$ of key distribution there exists an enveloping model
$\beta$ that agrees with $\alpha$ for commands dealt with by $\alpha$,
but encompasses other possibilities, and leads to conclusions about
security that conflict with those implied by model $\alpha$.  By
drawing on the quantum-mechanical separation of states and outcomes,
this proof used more than was really necessary.  All that is necessary
is the separation made in quantum mechanics between what happens at an
occurrence of a measure, {\em an} outcome, and what might have
happened, expressed as the {\em set of possible outcomes}.  This
separation is found not only in quantum mechanics, but in any
statistical theory, and in particular in the usual electrical
engineering of ``non-quantum'' systems, based on Maxwell's
electromagnetics to which one adjoins ideas of noise or the generation
of random signals.  In this non-quantum framework, the statistical
outlook alone allows one to introduce CPC's as a medium in which to
see a divide between models and instruments within the CPC's that
manage both.  Doing this, one puts the statements of a model in the
form $\Pr_\alpha(j|b_A,b_E)$.  Then Propositions 1 and 2 can be proved
without resort to quantum mechanics, so again the issues considered
above arise: (a) what else might Eve measure that a model used by
Alice and Bob has failed to account for, and (b) how might clock
pumping help Alice and Bob?  Thus the uncloseable possibility of
enveloping any model $\alpha$ by another model $\beta$ that expresses
extra conditions of the use of instruments is no peculiarity of using
quantum rather than non-quantum models; it is endemic to any
cryptographic modeling that invokes probabilities.

\acknowledgments 
We thank Howard E. Brandt for reading
an early draft and giving us an astute critique, indispensable to this
paper.  The situations described in which Diana, Alice, Bob, and
Eve take part are what Wittgenstein called {\em language games}
\cite{witt1}, with the language being the quantum mechanics of
CPC-oriented models.

\vfil
\eject 
\appendix
\section{Leakage channel in case Eve uses a probe}

This appendix proves for the case of Eve using a probe the analog of
Proposition 1: if one model with its inner products predicts a set of
probabilities for outcomes, so does another model having smaller inner
products.  (Thus, as in the segmented case, inner products depend on a
choice of model undetermined by measured data.) The proof here makes
no requirement that Alice choose among only two state vectors; she can
choose from a set of any size. Transposing to the CPC-context a model
expressing Eve's use of a probe \cite{slutskyPRA}, one obtains a model
$\alpha$ that assumes:
\begin{enumerate}
\item a set of Alice's commands $A_\alpha$; 
\item a Hilbert space ${\cal H}_{{\rm sig},\alpha}$ for Alice's
signals and a disjoint Hilbert space ${\cal H}_{{\rm
probe},\alpha}$ for Eve's probe;
\item a function assigning Alice's states to commands (here
only Alice's commands), $|v_\alpha \rangle : A_\alpha \rightarrow
{\cal H}_{{\rm sig},\alpha}$; 
\item a fixed starting state for Eve's probe of $|e_\alpha \rangle \in
{\cal H}_{{\rm probe},\alpha}$; 
\item a set $E_\alpha$ of Eve's possible commands to her measuring
instruments;
\item unitary operators $U_\alpha(b_E)$ for $b_E \in E_\alpha$ acting 
on the product Hilbert space ${\cal H}_{{\rm probe},\alpha} \otimes
{\cal H}_{{\rm sig},\alpha}$ for the interaction of Eve's probe
with Alice's signal;
\item a set $O_E$ of possible outcomes of Eve's measurements, indexed by
$j_E$;
\item for each of Eve's commands $b_E \in E_\alpha$, a POVM
$M_{E,\alpha}(b_E)$ with detection operators $M_{E,\alpha}(b_E;j_E)$
that act on ${\cal H}_{{\rm probe},\alpha}$;
\item a POVM for Bob's receiver acting on ${\cal H}_{{\rm sig},\alpha}$,
with possible outcomes indexed by $j_B$ and detection operators
$M_B(j_B)$.
 \end{enumerate} This produces a quantum
mechanical model $\alpha$ in which
\begin{eqnarray} \lefteqn{\mbox{$\Pr_\alpha$}(j_B,j_E|b_A,b_E) 
}\quad\nonumber \\ &=& 
\langle v_\alpha (b_A)|\langle e_\alpha |
U_\alpha^{\dagger}(b_E)  [M_B(j_B)\otimes M_E(b_E;j_E)] U_\alpha(b_E)
|e_\alpha |\rangle  |v_\alpha(b_A) \rangle.  
\end{eqnarray}

\begin{prop} Given any such (probe) model $\alpha$ with inner products
$\langle v_\alpha(b_A)|v_\alpha(b'_A)\rangle$ and any $0 \leq r <
1$, there is a model $\beta$ that gives the same conditional probabilities
of Eve's and Bob's outcomes for each command $b_E \in E_\alpha$ and for all
of Bob's commands, so
\begin{equation} (\forall
b_A \in A_\alpha, b_E \in E_\alpha) \mbox{$\Pr_\beta(j_E|b_A,b_E)$} =
\mbox{$\Pr_\alpha(j_E|b_A,b_E)$} \label{eq:same2} \end{equation} while
\begin{equation} (\forall b_A \neq b'_A) |\langle
v_\beta(b_A)|v_\beta(b'_A) \rangle | = r
|\langle v_\alpha(b_A)|v_\alpha(b'_A)\rangle | .
\label{eq:p_alpha2} \end{equation} \end{prop}

\noindent{\em Proof}:  The proof extends the construction used
in the proof of Proposition 1, with a model $\beta$ defined by
\begin{enumerate} 
\item the same command set for Alice: $A_\beta = A_\alpha$;
\item signals expressed by a vector intended by Alice, as in model
$\alpha$, tensored in to an unintended vector in an additional Hilbert
space ${\cal H}_{\rm leak}$, so Alice
produces vectors $|v_\beta(b_A) \rangle = |w_\beta(b_A) \rangle
\otimes |v_\alpha(b_A) \rangle$, with $|w_\beta(b_A) \rangle \in
{\cal H}_{\rm leak}$; 
\item a fixed starting state for Eve's probe of $|e_\alpha \rangle \in
{\cal H}_{{\rm probe},\alpha};$ 
\item a larger set of commands for Eve, $E_\beta = E_\alpha
\sqcup E_{\rm extra}$ (disjoint union);
\item unitary operators $U_\beta(b_E)$ acting on ${\cal H}_{\rm leak}
\otimes {\cal H}_{{\rm probe},\alpha} \otimes 
{\cal H}_{{\rm sig},\alpha}$ for the interaction of Eve's probe
with Alice's signal, defined so that \begin{equation}
U_\beta(b_E) = \left\{ \begin{array}{l}
{\bf 1}_{\rm leak} \otimes U_\alpha(b_E)
\mbox{ for all } b_E \in E_\alpha, \\[10pt]
\mbox{Eve's choice of unitary } U_\beta(b_E) \mbox{ if } b_E \in E_{\rm
extra}; \end{array} \right. \end{equation}
\item a POVM-valued function of Eve's commands to her measuring
instruments, with detection operators defined so
\begin{equation} M_\beta(b_E;j_E) = \left\{ \begin{array}{l} {\bf 1}
\otimes M_\alpha(b_E;j_E) \mbox{ for all } b_E \in E_\alpha, \\ 
\mbox{Eve's choice of POVM on } \\ {\cal H}_{\rm leak} \otimes
{\cal H}_{{\rm probe},\alpha} \mbox{ if } b_E \in E_{\rm
extra}. \end{array} \right. \end{equation} \end{enumerate} 
According to model $\beta$, if Eve chooses any measurement command of
$E_\alpha$, Eq.\ (\ref{eq:same2}) holds.  But model $\beta$ speaks not
of the vectors $|v_\alpha(b_A) \rangle$ but of other vectors having an
inner product (relevant to the security of quantum key distribution)
of \begin{equation} |\langle v_\beta(b_A)|v_\beta(b'_A) \rangle | =
|\langle w(b_A)| w(b'_A) \rangle | |\langle v_\alpha(b_A)
|v_\alpha(b'_A) \rangle|. \label{eq:winner2} \end{equation} The unit
vectors $|w(b_A) \rangle$ can be specified so that $|w(b_A) \rangle =
r^{1/2} |u(b_A) \rangle + (1-r) |u_0 \rangle$, with $\langle u(b_A)
|u(b'_A) \rangle = 0$ for all $b_A \neq b'_A$ and $\langle
u(b_A)|u_0 \rangle = 0$ for all $b_A \in A_\beta$.  With this
specification Eq.\ (\ref{eq:p_alpha}) holds, and furthermore $r$ can
be chosen as small as one wishes. $\Box$

\goodbreak

\end{document}